\documentclass[12pt]{article}
\usepackage[centertags]{amsmath}
\usepackage{amssymb}
\usepackage{graphicx}

\begin{document}

\title{Boolean Network Approach to Negative Feedback Loops of the p53 Pathways: Synchronized Dynamics and Stochastic Limit Cycles}

\author{Hao Ge\footnote{
Centre for Computational System Biology and Mathematical Department,
Fudan University, Shanghai 200433, P.R.China; email:
edmund\_ge@tom.com} \and Min Qian \footnote{School of Mathematical
Sciences, Peking University, Beijing 100871, P.R.China}
       }
\maketitle
\date{}

\begin{abstract}
Deterministic and stochastic Boolean network models are build for
the dynamics of negative feedback loops of the p53 pathways. It is
shown that the main function of the negative feedback in the p53
pathways is to keep p53 at a low steady state level, and each
sequence of protein states in the negative feedback loops, is
globally attracted to a closed cycle of the p53 dynamics after being
perturbed by outside signal (e.g. DNA damage). Our theoretical and
numerical studies show that both the biological stationary state and
the biological oscillation after being perturbed are stable for a
wide range of noise level. Applying the mathematical circulation
theory of Markov chains, we investigate their stochastic
synchronized dynamics and by comparing the network dynamics of the
stochastic model with its corresponding deterministic network
counterpart, a dominant circulation in the stochastic model is the
natural generalization of the deterministic limit cycle in the
deterministic system. Moreover, the period of the main peak in the
power spectrum, which is in common use to characterize the
synchronized dynamics, perfectly corresponds to the number of states
in the main cycle with dominant circulation. Such a large separation
in the magnitude of the circulations, between a dominant, main cycle
and the rest, gives rise to the stochastic synchronization
phenomenon.
\begin{flushleft}
{\bf KEY WORDS:} Hopfield network; Stochastic Boolean network;
Boltzmann machine; circulation; stochastic limit cycles;
synchronization; p53 protein;
\end{flushleft}
\end{abstract}

\section{Introduction}

    Thousands of papers have been reported in the field of p53
protein during the last three decades since 1979, including both the
experimental and theoretical analysis. p53, the tumor suppressor,
transcriptionally activates Mdm2, which in turn targets p53 for
degradation (Piette et al. 1997, Prives et al. 1998, Vogelstein et
al. 2000, Momand et al. 2000, Michael et al. 2003), keeping p53 at a
low steady state level. The low steady state of p53 is essential for
normal cell proliferation because p53 induces either cell cycle
arrest or programmed cell death. The concentration of p53 increases
in response to stress signals, such as DNA damage, inducing a
transition to oscillations of p53 level. Following stress signal,
p53 also activates transcription of several hundred genes, which are
involved in the program of cell cycle arrest, apoptosis, cellular
senescence and DNA repair. Therefore, it is important to note that
many additional proteins interact with p53 and Mdm2, so that a
number of positive and negative autoregulatory feedback loops acting
upon the p53 response are embedded inside a network with a few
additional interactions (Harris et al. 2005).

    Several mathematical models have been proposed to explain the damped
oscillations of p53, either in cell population or in a single-cell,
most of which are deterministic model of ordinary differential
equations (Mihalas et al. 2000, Lev Bar-Or et al. 2000, Monk et al.
2003, Ma et al. 2003, Ciliberto et al. 2005). On the other hand, the
stochastic nature of p53 dynamics has also been observed and caused
more and more interest nowadays (Ma et al. 2003).

    However, in many cellular biochemical modelling, a detailed
model, whether the deterministic one based on mass-action law,
Michaelis-Menten kinetics and Hill function, or the stochastic
molecular number-based CME({\bf chemical master equation}), is not
warranted because of a lack of enough quantitative experimental
data. Alternatively, one can develop discrete state network model,
i.e., deterministic and stochastic Boolean networks, of a complex
biological system using the available information on the activation
and repression from one signaling molecules to another, e.g., the
kind of signaling wiring diagram as the Fig. 1 in (Li et al. 2004),
where Li et. al. have developed a discrete Boolean model by applying
the approach of Hopfield for neural networks for yeast cell-cycle
regulatory network with 11 nodes. The main results of (Li et al.
2004) are that the network is both dynamically and structurally
stable. The biological steady state, known as the G1 phase of a cell
cycle, is a global attractor of the dynamics; the biological
pathway, i.e., the returning to G1 phase after perturbation, is a
globally attracting dynamic trajectory. The deterministic Boolean
model has been further extended to incorporate stochastic dynamics
(Zhang et al.2006). They found that both the biological steady state
and the biological pathway are well preserved under a wide range of
noise level. Furthermore, recently we investigated the synchronized
dynamics and nonequilibrium steady states in the stochastic yeast
cell-cycle network (Ge et al. 2007), applying the mathematical
circulation theory of Markov chains(Jiang et al. 2004).

In the present paper, deterministic and stochastic Boolean network
models are build for the dynamics of negative feedback loops of the
p53 pathways. It is shown that the main function of the negative
feedback in the p53 pathways is to keep p53 at a low steady state
level, and each sequence of protein states in the negative feedback
loops, is globally attracted to a closed cycle of the p53 dynamics
after being perturbed by outside signal (e.g. DNA damage). Our
theoretical and numerical studies show that both the biological
stationary state and the biological oscillation after being
perturbed are stable for a wide range of noise level. Applying the
mathematical circulation theory of Markov chains, we investigate
their stochastic synchronized dynamics and by comparing the network
dynamics of the stochastic model with its corresponding
deterministic network counterpart, a dominant circulation in the
stochastic model is seen to be the natural generalization of the
deterministic limit cycle in the deterministic system.

It is shown that a large separation in the magnitude of the
circulation, between a dominant main cycle and the rest, gives rise
to the stochastic synchronization phenomenon and the stochastic
global attractive behavior, and moreover the power spectrum of the
trajectory has a main peak, whose period converges just to the
number of states in the dominant cycle. Furthermore, the net
circulation of the dominant cycle increases monotonically with the
noise-strength parameter $\beta$, approaching its deterministic
limit; while the circulation of all the other cycles approaches zero
very fast when $\beta$ is quite large. Together, these observations
provide a clear picture of the nature of the synchronization and
stochastic limit cycles in a stochastic network in terms of the
probabilistic circulation of NESS (nonequilibrium steady states)
(Jiang et al. 2004).

For the completeness of the work, in supporting information, we give
a theoretical sketch of some relevant results on biological
networks, including a classification of the deterministic and
stochastic Boolean networks and their correspondence. Also a short
introduction of the mathematical theory of stochastic circulation
for Markov chains is introduced and applied to deterministic and
stochastic networks. It is shown that the stochastic Boolean network
is reversible if and only if the matrix $T$ in the model is
symmetric, and the net NESS circulation is strictly positive as long
as the probability of the directed cycle is larger than that of its
reversal.

\section{Boolean network approach}

Since the influential work of J.J. Hopfield in 1980s' (Hopfield
1982, Hopfield 1984), the deterministic Boolean (Hopfield) network
has been applied to various fields of sciences. Amit (Amit 1989) has
introduced a temperature-like parameter $\beta$ that characterizes
the noise in the network and constructed a probabilistic Boolean
network called Boltzmann machine. The deterministic and stochastic
Boolean networks have found wide range of applications in biological
networks.

In our model, the states of the nodes(proteins, DNAs or RNAs) in the
network at the n-th step are represented by variables
$X_n=(X_n^1,X_n^2,\cdots,X_n^N)$ respectively, where $N$ is the
number of nodes in the network. Each node $i$ has only two values,
$X_n^i=1$ and $X_n^i=-1$, representing the active state and the
reset state of this node respectively.

{\bf Deterministic Boolean network model:}

The deterministic model consider in the present paper is a
deformation of model $A1$ in supporting information. Let us suppose
$N$ is a fixed integer, $S=\{1,2,\cdots,N\}$. We take the state
space as $\{-1,1\}^S$. Denote the state of the $n$-th step as
$X_n=(X_n^1,X_n^2,\cdots,X_n^N)$, then the dynamic is as follows:

If $X_n\neq(-1,-1,\cdots,-1)$, then
\begin{eqnarray}
    X_{n+1}^i=\left\{\begin{array}{ll}\textrm{sign}
    (H_i),&\textrm{if}~\sum_{j=1}^N T_{ij}X_n^j\neq U_i;\\X_n^i,&\textrm{if}~\sum_{j=1}^N T_{ij}X_n^j=U_i;\end{array}\right.
\end{eqnarray}
where the function
$\textrm{sign}(x)=\left\{\begin{array}{ll}1&x>0;\\0&x=0;\\-1&x<0,\end{array}\right.$
$H_i=\sum_{j=1}^N T_{ij}X_n^j-U_i$ is the input to the i-th node and
$U_i~(1\le i\le N)$, given {\it a priori}, are called the threshold
of the $i$-th unit.

And if $X_n=(-1,-1,\cdots,-1)$, then $X_{n+1}=X_n$ when the
parameter $\gamma=0$ representing that the system is under normal
environment, and $X_{n+1}=(1,-1,-1,\cdots,-1)$ when the parameter
$\gamma=1$ representing there are some perturbation(signal) of the
system (e.g. DNA damage) emerge.

{\bf Stochastic Boolean Network model: }

The stochastic model consider in the present paper is a deformation
of model $C$ in supporting information, which is similar to the
model in (Ge et al. 2007). Consider a Markov chain
$\{X_n=(X_n^1,X_n^2,\cdots,X_n^N), n=0,1,2,\cdots\}$ on the state
space $\{-1,1\}^S$, with transition probability given as follows:
\begin{equation}
    P(X_{n+1}|X_n)=\prod_{i=1}^N P_i(X_{n+1}^i|X_n),\label{Eq_pd}
\end{equation}
where
$$P_i(X_{n+1}^i|X_n)=\frac{\exp(\beta X_{n+1}^iH_i)}
    {\exp(\beta H_i)+\exp(-\beta H_i)},$$
in which $H_i=\sum_{j=1}^N T_{ij}X_n^j-U_i$ is the input to the i-th
node, if $\sum_{j=1}^N T_{ij}X_n^j\neq U_i$ and
$X_n\neq(-1,-1,\cdots,-1)$ or $i\geq 2$,
$$P_i(X_{n+1}^i|X_n)=\left\{\begin{array}{ll}\frac{1}{1+e^{-\alpha}},&X_{n+1}^i=X_n^i;\\
\frac{e^{-\alpha}}{1+e^{-\alpha}},&X_{n+1}^i=1-X_n^i,\end{array}\right.$$
if $\sum_{j=1}^N T_{ij}X_n^j=U_i$ and $X_n\neq(-1,-1,\cdots,-1)$ or
$i\geq 2$, and
$$P_1(X_{n+1}^i|X_n)=\left\{\begin{array}{ll}1-\gamma,&X_{n+1}^i=X_n^i,\\
\gamma,&X_{n+1}^i=1-X_n^i,\end{array}\right.$$ if
$X_n=(-1,-1,\cdots,-1)$, where the parameter $0\leq\gamma\leq 1$
still represents the stochastic perturbations of the system from
extracellular environment.

In the above equation, $\alpha(>0)$, $\beta(>0)$, $T_{ij}$ and $U_i~
(1\le i,j\le N)$ are parameters of the model. The positive
temperature-like parameter $\beta$ represents noise in the system
from the perspective of statistical physics (Amit 1989, Zhang et al.
2006). Noticeably, the actual noises within a cell might not be
constant everywhere, but here we use a system-wide noise measure for
simplicity. To characterize the stochasticity when the input $H_i$
to a node is zero, we have to introduce another parameter $\alpha$.
This parameter controls the likelihood for a protein to maintain its
state when there is no input to it.

The previous parameters $\beta$ and $\alpha$ represent the
intracellular noise due to thermodynamic fluctuations, and on the
other hand, we need another parameter $\gamma$ to characterize the
extracellular signal strength with appropriate stochasticity. Since
the signal is not purely disordered, we can not express the
probability $P_1(X_{n+1}^i|X_n)$ when $X_n=(-1,-1,\cdots,-1)$ as the
exponential form analogous to the Boltzmann distribution, hence we
directly regard $\gamma$ as the probability transiting from
$(-1,-1,\cdots,-1)$ to $(1,-1,\cdots,-1)$.

In our models below, $T_{ij}=1$ for a arrow from protein $j$ to
protein $i$, and $T_{ij}=-1$ for a horizontal bar instead of
arrowhead from protein $j$ to protein $i$(Fig. \ref{fig0},
\ref{fig7}). And it is indispensable to point out that
self-connections haven't been taken into consideration in our models
for simplicity.

Similar to Proposition II.6 in supporting information, with the same
initial distribution, when $\gamma\rightarrow 0$, and $\alpha$,
$\beta\rightarrow\infty$, then the stochastic Boolean network model
converges to the corresponding deterministic Boolean network model
with parameter $\gamma=0$; and when $\gamma\rightarrow 1$, and
$\alpha$, $\beta\rightarrow\infty$, then the stochastic Boolean
network model converges to the corresponding deterministic Boolean
network model with parameter $\gamma=1$.

\section{Steady states and synchronized dynamics of the p53 pathways}

    From biochemical perspective, the microscopic variables
for a cellular regulatory network are the concentrations, or
numbers, of various mRNAs, regulatory proteins, and cofactors. If
all the biochemistry were known, then the dynamics of such a network
would be represented by a chemical master equation (McQuarrie 1967,
Gillespie 1977). Unfortunately, much of the required information is
not available, nor such a ``fully-detailed'' model will always be
useful. Phenomenologically the concentrations of key players of a
biochemical regulatory network can often be reduced to two or three
states, such as resting state, activated state, inactivated state,
etc. (Fall et al. 2002).  The interactions between these states are
usually determined from experimental data.

The present study will build simple deterministic and stochastic
Boolean network models for several negative feedback loops of the
p53 pathways, and more important, provides a sound mathematical
explanation of the synchronized dynamics and stochastic limit cycles
in the stochastic Boolean network model after being perturbed by
stress signals, in terms of the theory of nonequilibrium
circulations (Jiang et al. 2004). This makes the description of
pathway more definite and penetrating.

\subsection{The core regulation}

Negative feedback loops, composed of one transcription arm and one
protein-interaction arm, are a common network motif across
organisms. p53, the tumor suppressor, transcriptionally activates
Mdm2, which in turn targets p53 for degradation. Although it is
believed that the existence of negative feedback loop in biological
systems can always gives rise to oscillations, yet it is not
sufficient for the appearing of a limit cycle in the deterministic
ODE model from the mathematical point of view.

\begin{figure}[h]
\centerline{\includegraphics[width=2in,height=1.5in,angle=270]{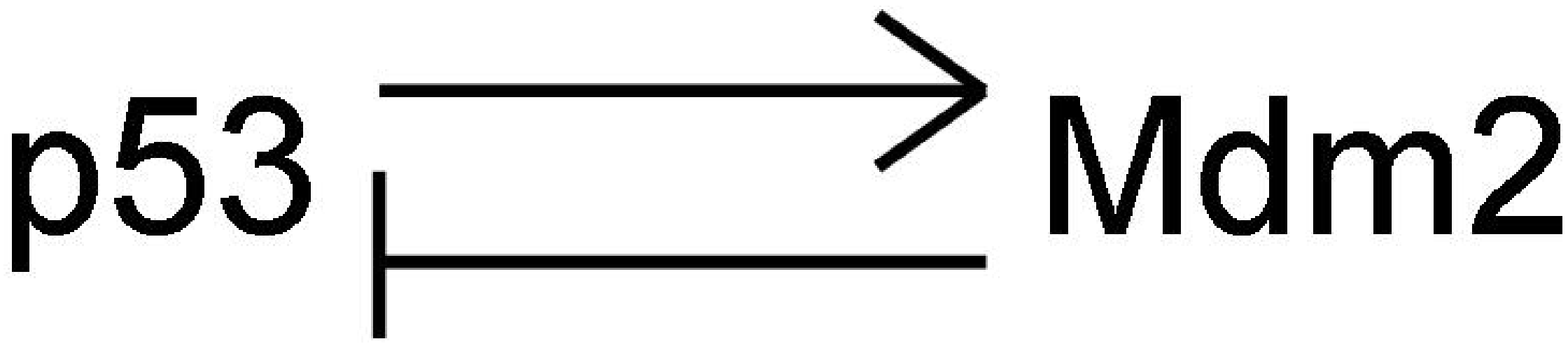}}
\caption[fig0]{Negative feedback loop of p53 and Mdm2. Arrows denote
stimulatory interactions, whereas horizontal bars instead of
arrowheads indicate inhibitory influences.} \label{fig0}
\end{figure}

A bit out of expectation is, if we use the deterministic Boolean
network even to the simplest case(Fig.\ref{fig0}), the existence of
negative feedback already gives rise to a limit cycle corresponding
to oscillations in a biological system, when there exists the
outside signal (i.e. $\gamma=1$).

\marginpar{Fig 1.}

{\bf Model:} From Section II(For Fig.\ref{fig0}).

The first node $X_n^1$: p53; and the second node $X_n^2$: Mdm2.

$N=2$ is the number of nodes in the model. For simplicity, we set
thresholds as $U_1=U_2=0$. And the interacting matrix
$$T=\left[\begin{array}{ll}0&-1\\1&0\end{array}\right]$$

The deterministic model with $\gamma=0$ has a global attractor
$(-1,-1)$, which corresponds to the fact that the main function of
the negative feedback between $p53$ and $Mdm2$ is to keep $p53$ at a
low steady state level in normal cells.

On the other hand when $\gamma=1$, the deterministic model  has a
unique limit cycle consist of $4$ state, which is described by Table
\ref{tab0}. It corresponds to the fact that the stress signal (e.g.
DNA damage) will activate the protein p53(i.e. time-2 point in Table
\ref{tab0}) and induce a transition to oscillations of p53 level
after being perturbed from the outside environment.

\begin{table}[h]
\begin{center}
\begin{tabular}{crr}
Time&p53&Mdm2\\
\hline
1&1&-1\\
2&1&1\\
3&-1&1\\
4&-1&-1
\end{tabular}
\end{center}
\caption[tab0]{Cycle evolution of the protein states in the negative
feedback loop of p53 and Mdm2 after being perturbed.} \label{tab0}
\end{table}

Moreover, as long as the stress signal strength $\gamma$ is
sufficiently low $(0<\gamma<<1)$ or sufficiently high
$(0<1-\gamma<<1)$, the stochastic Boolean network model would
preserve the dynamics of the corresponding deterministic model at a
certain low level of noise. As it illustrates the same phenomenon as
the Cyclin G/Mdm-2 loop given below, so we won't show the data here.

Similar model can also be built and analyzed exactly by the same
reasoning, to other ubiquitin ligases that promote p53
ubiquitination and subsequent proteasomal degradation (Fig. 10 in
Harris et al. 2005). We only use the simplest example as a start,
and pass directly to a really delicate interesting case.

\subsection{Cyclin G/Mdm2 loop}

\marginpar{Fig 2.}

{\bf Model:} From Section II.

The first node $X_n^1$: p53; The second node $X_n^2$: cyclin G; The
third node $X_n^3$: PP2A cyclin G; The last node $X_n^4$: Mdm-2.

\begin{figure}[h]
\centerline{\includegraphics[width=2in,height=1.5in,angle=270]{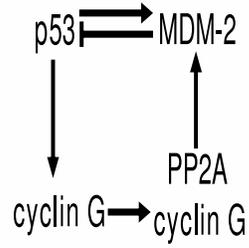}}
\caption[fig7]{Cyclin G/Mdm2 loop, redrawn from (Fig. 8 in Harris et
al. 2005).} \label{fig7}
\end{figure}

N=4 is the number of nodes in the model. For simplicity, we set
thresholds as $U_1=U_2=U_3=U_4=0$. And from Fig.\ref{fig7}, the
interacting matrix
$$T=\left[\begin{array}{llll}0&0&0&-1\\1&0&0&0\\0&1&0&0\\1&0&1&0\end{array}\right].$$

The deterministic model with $\gamma=0$ has a global attractor
$(-1,-1,-1,-1)$, which corresponds to the fact that $p53$ is kept at
a low steady state level in normal cells.

On the other hand, the deterministic model when $\gamma=1$ has a
unique limit cycle consist of $8$ state, which is described by Table
\ref{tab7}. It corresponds to the fact that the stress signal (e.g.
DNA damage) will activate the protein p53(i.e. time-2 point in Table
\ref{tab7}) and induce a transition to oscillations of p53 level
after being perturbed from the outside environment, which roughly
corresponds to the p53 pathway or biological trajectory described in
(Harris et al. 2005) ``One of the most active of the p53-responsive
genes is the cyclin G gene. It is rapidly transcribed to high levels
after p53 activation in a wide variety of cell types.''(i.e. Time-3
point in Table \ref{tab7}) ``The cyclin G protein makes a complex
with the PP2A phosphatase, which removes a phosphate residue from
Mdm-2, which is added to the Mdm-2 protein by a cdk kinase'' (i.e.
Time-4,5 points in Table \ref{tab7}). Our language in the present
paper is more definite and penetrating than the quotations.

\begin{table}[h]
\begin{center}
\begin{tabular}{crrrr}
Time&p53&Cyclin G&PP2A Cyclin G&Mdm-2\\
\hline
1&(-1&-1&-1&-1)\\
2&(1&-1&-1&-1)\\
3&(1&1&-1&-1)\\
4&(1&1&1&-1)\\
5&(1&1&1&1)\\
6&(-1&1&1&1)\\
7&(-1&-1&1&1)\\
8&(-1&-1&-1&1)
\end{tabular}
\end{center}
\caption[tab1]{Cycle evolution of the protein states in the Cyclin
G/Mdm-2 loop after being perturbed.} \label{tab7}
\end{table}

{\bf Numerical Simulation of the stochastic Boolean network model}

The numerical results about the cyclic motion of this stochastic
model  are quite similar to the stochastic Boolean network model of
the cell-cycle, which we have recently discussed (Ge et al. 2007).
The main conclusion is that, given the structure of the negative
feedback present in Fig. \ref{fig7}, the stochastic model still
exhibits well pronounced oscillations after being seriously
perturbed, which is excellently characterized by the circulation
theory introduced in the supporting information.

Numerical computations for the current model(Fig.\ref{fig7}), are
carried out with the famous Gillespie's method (Gillespie 1977) of
the stochastic Boolean network model using MATLAB, and the results
are given in the following figures. The network with $4$ binary
nodes has a total of $2^4=16$ number of states. Here, we can present
the dynamics of the network in terms of the integer states
$0,1,2,\cdots,2^4-1=15$ on a line. This 1-d system is reversible if
and only if the 4-d system is reversible.

\begin{figure}[h]
\centerline{\includegraphics[width=3in,height=2in]{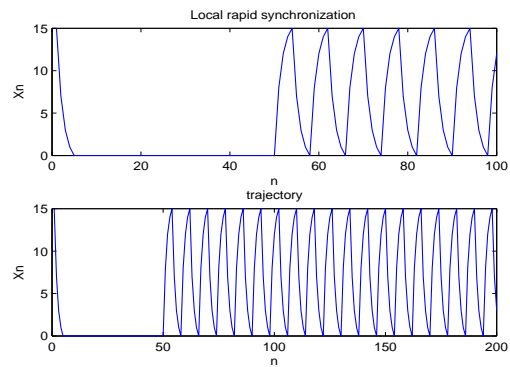}}
\caption[fig71]{Stochastic trajectory and synchronization of the
Cyclin G/Mdm-2 loop. Simulations are carried out with the parameters
$\alpha=5$, $\beta=5$ and the values of $\gamma$ transits from
$0.001$ to $0.999$ at time $n=50$.} \label{fig71}
\end{figure}

Fig. \ref{fig71} are the basic behavior of a random trajectory. The
upper panel shows that there arises the phenomenon of local rapid
synchronization like that observed in (Hopfield 1995) during a very
short time period after the value of $\gamma$ transits from $0.01$
to $0.99$ at time $n=50$, when $\beta$ and $\alpha$ are sufficiently
large. The lower panel is a random trajectory over a longer time.
Little deviation is shown from the deterministic trajectory in Table
\ref{tab7} after being perturbed at time $n=50$, which implies that
the stochastic model still leads to well pronounced oscillations
when the perturbation from the environment is sufficiently high.

\marginpar{Fig 3.}

\begin{figure}[h]
\centerline{\includegraphics[width=3in,height=2in]{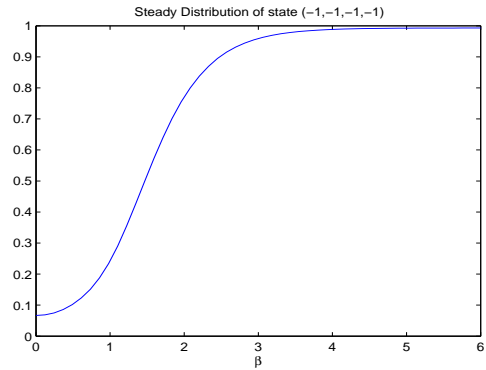}}
\caption[fig711]{Stationary distribution of the state
$(-1,-1,-1,-1)$ in the stochastic model. Simulations are carried out
with the parameters $\alpha=5$ and $\gamma=0.001$.} \label{fig711}
\end{figure}

Fig. \ref{fig711} shows the stationary distribution of the state
$(-1,-1,-1,-1)$ in the stochastic model which increasingly
approaches $1$ when $\beta$ tends to infinity. This excellently
corresponds well to the dynamics of the deterministic model when
$\gamma=0$. At large $\beta$ (low ¡°temperature¡± or small noise
level), the low level state $(-1,-1,-1,-1)$ is the most probable
state of the system. So analogous to the concept of the
deterministic model, this state $(-1,-1,-1,-1)$ can be regarded as
the global attractor of the stochastic model. Moreover, one observes
a phase-transition like behavior of the stationary distribution of
the state $(-1,-1,-1,-1)$ while varying the noise level $\beta$
(similar behavior has been seen in (Qu et al. 2002)).

\marginpar{Fig 4.}

Then we turn to investigate the synchronized dynamics when the
signal parameter $\gamma$ is sufficiently high(i.e.
$0<1-\gamma<<1$). The keys to understand synchronization behavior in
stochastic Boolean network models are ($i$) establishing a
correspondence between a stochastic dynamics and its deterministic
counterpart; and ($ii$) identifying the cyclic motion in the
stochastic models.

As there is a growing awareness and interest in studying the effects
of noise in biological networks, it becomes more and more important
to quantitatively characterize the synchronized dynamics
mathematically in stochastic models, because the concepts of limit
cycle and fixed phase difference no longer holds in this case.
Instead, physicists and biologists always have to characterize
synchronized dynamics by the distinct peak of some spectrum or just
only by observing the stochastic trajectories, which however may
cause ambiguities in the conclusion. Therefore, a logical
generalization of limit cycle to the stochastic model is well worth
to be developed.

In case of the stochastic models of biological networks, there does
exist a rather complete mathematical theory for the cyclic motion of
the corresponding Markov chains, which has been developed for more
than twenty years (Jiang et al. 2004, Kalpazidou 1995). One of the
most important concepts in this mathematical theory of NESS is the
circulation, which corresponds to the cycle kinetics in open
chemical systems (Hill 1989, Qian 2005). The details of mathematical
theory is supplied in the supporting information.

\begin{figure}[h]
\centerline{\includegraphics[width=3in,height=2in]{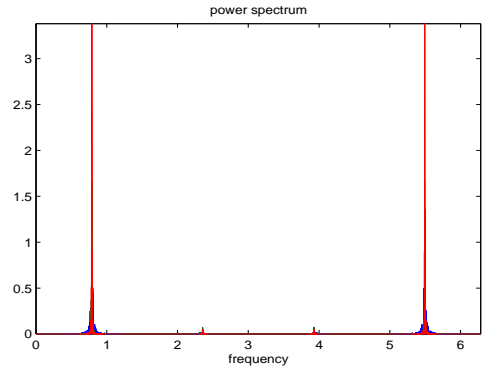}}
\caption[fig72]{Power spectrum of the overall trajectory in the
Cyclin G/Mdm-2 loop with the parameter $\alpha=5$ and $\gamma=0.999$
fixed, and with different $\beta$: blue: $\beta$=2.4; green:
$\beta$=4.8; red: $\beta=6$. The discrete Fourier transform causes
an alias; hence the spectrum is symmetric with respect to $\pi$ on
the $[0,2\pi]$ interval. } \label{fig72}
\end{figure}

To further characterize the synchronized dynamics, we give Fig.
\ref{fig72} which shows the Fourier power spectrum of the stochastic
trajectory with different values of $\beta$ respectively. Using
MATLAB, the discrete Fourier transform for time series
$\{x_1,x_2,\cdots,x_n\}$ is defined as
\begin{equation}
    y_m=\left|\sum_{k=1}^n x_ke^{-i(2\pi/n)(m-1)(k-1)}\right|,
    \hspace{0.4cm}
    \left(\frac{m-1}{n}2\pi,\ 1\le m\le n\right).
\end{equation}
Therefore, by the Herglotz theorem (Qian et al. 1997 p. 331), the
power spectrum of discrete trajectory has a symmetry
$y_m=y_{n+2-m}$. For different sets of parameters, we found all the
calculations give the same outstanding main peak in the Fig.
\ref{fig72}. It is important to mention here that by ergodicity,
different trajectories give the same power spectrum for any $\beta$
that is sufficiently large.

\marginpar{Fig 5.}

The single dominant peak in Fig. \ref{fig72} implies there exists a
global synchronization and a globally attractive phenomenon. Note
that by representating our maps, one-to-one, from the $N$ binary
nodes to the integers $0-15$, the synchronized behavior is
preserved. It is possible that the map will cause some distortion in
the power spectrum.

\begin{figure}[h]
\centerline{\includegraphics[width=3in,height=2in]{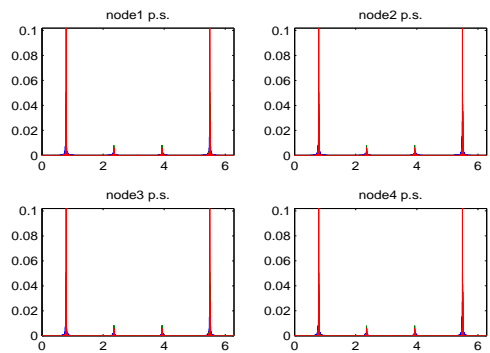}}
\caption[fig73]{Power spectra of individual nodes in the Cyclin
G/Mdm-2 loop show a synchronization among all the nodes with the
parameter $\alpha=5$ and $\gamma=0.999$ fixed, and different
$\beta$: blue: $\beta$=2.4; green: $\beta$=4.8; red: $\beta$=6.}
\label{fig73}
\end{figure}

To further illustrate the synchronized behavior, Fig. \ref{fig73}
shows the power spectra of all the 4 individual nodes in the
network. While subtle details are different, all exhibit the
dominant peak, similar to that of the overall dynamics. This
demonstrates further that synchronized dynamics is presented in the
network.

\marginpar{Fig 6.}

\begin{figure}[h]
\centerline{\includegraphics[width=3in,height=2in]{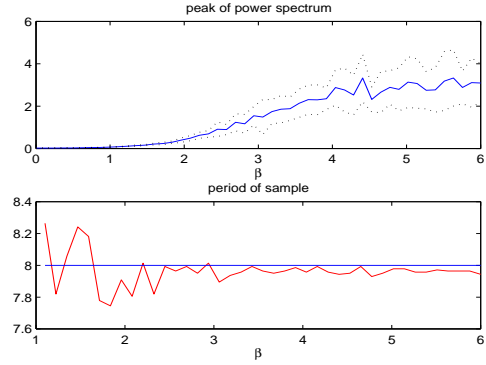}}
\caption[fig74]{Magnitude and period of the dominant power-spectral
peak as functions of $\beta$ in the Cyclin G/Mdm-2 loop, with
$\alpha=5$ and $\gamma=0.999$. In the upper panel, the magnitude of
the dominant power-spectral peak is averaged over 20 simulations.
The solid curve is the mean, and the dotted curves are the mean
$\pm$ standard deviation. The period approaches to 6 (the horizonal
line) with increasing $\beta$.} \label{fig74}
\end{figure}

Fig. \ref{fig74} plots the magnitude and the period of the dominant
peak of the power spectrum in Fig. \ref{fig72} respectively, as
functions of the noise strength, i.e., the parameter $\beta$. It
shows, as we have predicted, that the period converges to 8 which
corresponds perfectly to the number of states in the main cycle of
Table \ref{tab7} when $\beta$ is large. We also put error bars on
the upper panel of Fig. \ref{fig74} with various values of $\beta$.

%There seems to be a surprising appearance of period$=2$ cycles for
%$\beta=4.3$ and $\beta=5.6$. We believe that these dramatic bumps in
%the red curve are due to the technical misleading rather than a
%really important phenomenon, since the numerical calculation is
%based on the simulated stochastic trajectories and we only pick the
%power-spectral peak with highest magnitude. It is found that the
%values of $\beta$ at which the dominant power-spectral peak is of
%frequency $\pi$ (period=2) are quite different according to
%different simulated trajectory samples, but the period still
%preserves at $6$ when $\beta$ and $\alpha$ are sufficient large.
%Actually, when $\beta=4.3$ or $\beta=5.6$ of the simulation results
%shown in Fig. \ref{fig14}, the power spectrum has two strong peaks
%at frequencies $\frac{\pi}{3}$(period=6) and $\pi$(period=2)
%respectively, but the former is only slightly lower than the
%latter(data not shown).

\marginpar{Fig 7.}

\begin{figure}[h]
\centerline{\includegraphics[width=3in,height=2in]{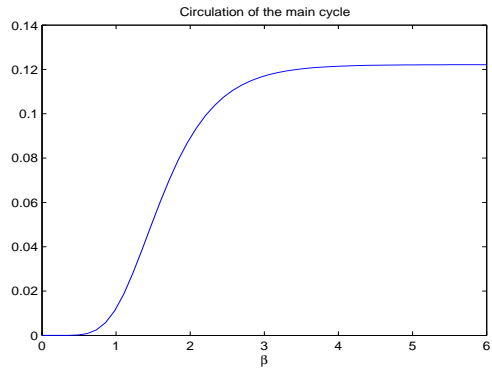}}
\caption[fig75]{Net circulation of the main cycle in the Cyclin
G/Mdm-2 loop as a function of the noise strength, $\beta$, with the
parameter $\alpha=5$ and $\gamma=0.999$ fixed.} \label{fig75}
\end{figure}

Finally, Fig. \ref{fig75} shows how the net circulation of the
dominant, main cycle varies with $\beta$, applying the determinant
presentation of circulations according to Theorem II.4 in the
supporting information. Note that circulation is just the
time-averaged number of appearance of certain cycle along the
stochastic trajectory of our model, and net circulation is just the
difference between the circulations along the positive direction and
negative direction of a specific cycle. It is clearly seen that the
net circulation of the main cycle increases monotonically to
$\frac{1}{8}$ that is just the reciprocal of the number of states in
the main cycle, which implies the appearing of more and more
distinct synchronization and global attractive behavior with
increasing $\beta$. The direction of the net circulation does not
change. It is also found that the circulation of negative direction
along the main cycle is always quite low compared to the circulation
of positive direction.

\marginpar{Fig 8.}

The net circulations of all the cycles are very small when $\beta$
and $\alpha$ are near zero, since the system is close to
equilibrium(reversible) state when $\alpha=\beta=0$ according to
lemma II.9 in supporting information.

\begin{figure}[h]
\centerline{\includegraphics[width=3in,height=2in]{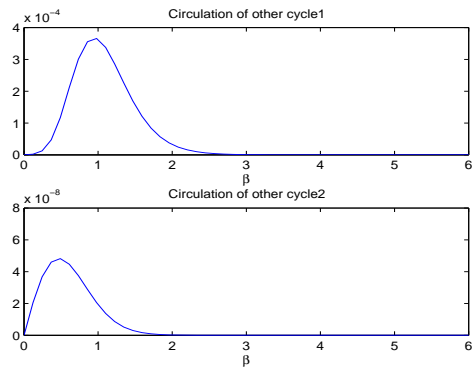}}
\caption[fig76]{Net circulation of several other cycles in the
Cyclin G/Mdm-2 loop as a function of the noise strength, $\beta$,
with the parameter $\alpha=5$ and $\gamma=0.999$ fixed.}
\label{fig76}
\end{figure}

For large $\beta$, the net circulations of all the cycles except the
main cycle are also very small by numerical simulation using the
determinant expression in Theorem II.4 of supporting information.
All the circulations of non-main cycles actually decrease with
increasing $\beta$ when $\beta$ is large. Examples are shown in Fig.
\ref{fig76}. This large separation in the magnitudes of the weights
gives rise to the stochastic synchronization, and this stochastic
limit cycle can be defined as a {\bf ``stable''} one, whose
attractive domain is global.

\marginpar{Fig 9.}

As in (Zhang et al. 2006, Ge et al. 2007), we also notice that there
exists an inflection point in the curve in Fig. \ref{fig75}. This
implies a cooperative transition of the net circulation of the main
cycle while varying the noise level $\beta$, which equivalently
means some sort of ``phase transition'' around $\beta=2$ from
chaotic fluctuations(period=infinity) at small $\beta$ to periodic
fluctuations (period=8) at large $\beta$. Certainly, the implication
of this observation remains to be further elucidated.

\section{Discussion}

{\bf From detailed models to simplified Boolean network approach}

Deterministic nonlinear mathematical models, based on the Law of
Mass Action, have been traditionally used for biochemical reaction
networks. Furthermore, noises are unavoidable in small biochemical
reaction systems such as those inside a single cell. Stochastic
models with chemical master equations (CME) (McQuarrie 1967, Van
Kampen 1981) should be developed, which has already provided
important insights and quantitative characterizations in some cases
of the biochemical system (Fox et al. 1994, Fox 1997, Zhou et al.
2005). Ref. (Wilkinson 2006) is a good introduction to the
stochastic modelling in biology.

In many cellular biochemical modelling, it is impossible to build a
detailed, molecular number-based CME model due to a lack of
quantitative experimental data. Thus, one often seeks a simplified
network model based on simple binary states of the signaling
molecules. This leads to the Boolean, Markov network model of the
p53 pathways we studied in the present paper.

{\bf Modelling the oscillatory dynamics of p53 pathways after being
perturbed}

The realization that p53 is a common denominator in human cancer has
stimulated an avalanche of research since 1989. The p53 gene can
integrate numerous signals that control cell life and death, and
 damped oscillations for p53 and Mdm2 has been observed and
modeled (Lev Bar-Or et al. 2000, Ma et al. 2003). The Boolean
network models built in the present paper, which omit some
parameters to represent the degradation of the p53 protein in cells,
only try to explain the mechanism of the oscillatory behavior of the
p53 dynamics after being perturbed by stress signals rather than
exhibiting the damped oscillations. The signal parameter $\gamma$
plays a very important role in the model similar to the cell cycle
model in (Ge et al. 2007), which can induce the level of p53 from a
low steady state to oscillated behaviors.

On the other hand, ideally, one should try to combine these Boolean
network models of negative feedback loops together to construct a
clear and integrated picture of the p53 pathways, maybe especially
including several positive feedback loops (Harris et al. 2005). But
unfortunately we haven't developed a reasonable way to do so.

{\bf Synchronization and circulation in stochastic Boolean network}

Synchronization is an important characteristics of many biological
networks (Strogatz 2003, Winfree 2000) whose dynamics has been
modelled traditionally by deterministic, coupled nonlinear ordinary
differential equations in terms of regulatory mechanisms and kinetic
parameters (Murray 2002, Fall et al. 2002). Two important classes of
biological networks which have attracted wide attentions in recent
years are neural networks (Scott 2002) and cellular biochemical
networks (Goldbeter 1996).

As we know, the occurrence of a deterministic limit cycle in an ODE
model is the hallmark of a synchronization phenomenon, while such a
definite concept no longer holds in a stochastic system. It is
observed that in our present model of p53 pathways, the trajectory
concentrates around a main cycle, which we call {\bf stochastic
limit cycle} and is very similar to in many aspects the limit cycle
of a deterministic model. Furthermore, the present work shows that
the circulation in an irreversible Markov chain (Jiang et al. 2004)
is a reasonable generalization of the concept of deterministic limit
cycles, and provides a sound mathematical explanation of the
synchronization in the stochastic Boolean network model.

{\bf Stability and robustness of the p53 genetic networks}

Biological functions in living cells are controlled by protein
interaction and genetic networks, and have to be robust to function
in complex (and noisy) environments. More robustness would also mean
being more evolvable, and thus more likely to survive.

More precisely, these molecular networks should be dynamically
stable against various fluctuations which are inevitable in the
living world. Therefore, the stability and robustness of Boolean
network model can not be determined if the noise hasn't been
introduced into the model, since it is not reasonable to simply
apply the Euclidean topology in mathematics to such a real
biological discrete model not to say the magnitudes of the different
attractive basins of fixed states and limit cycles as in the
deterministic models. For instance, the recent analysis of
stochastic cell-cycle Boolean networks (Ge et al. 2007, Zhang et al.
2006) just claimed the stability and robustness of the previous
deterministic model (Li et al. 2004), in which Li, et.al. announced
but didn't really investigate the robustness of this model against
perturbation. Hence, the numerical and theoretical studies in the
present paper in some sense excellently exhibit the stability and
robustness of the p53 genetic networks.

\section*{Acknowledgement}

The authors would like to thank Professor Minping Qian in Peking
University for calling our attention to the p53 network.

\end{document}

\title{Supporting Information}

\author{Hao Ge\footnote{
School of Mathematical Sciences, Peking University, Beijing
100871, P.R.C.; email: edmund\_ge@tom.com} \and Min Qian
\footnote{School of Mathematical Sciences, Peking University,
Beijing 100871, P.R.C.}
       }
\maketitle
\date{}

\section{Theoretical Sketch of Some Relevant Results on Deterministic and Stochastic Boolean Networks}

This section is mainly restated from (Ge et al. 2007).

   We first give a brief account of the deterministic Hopfield networks
and its stochastic incarnation, the stochastic Boolean networks
(sometimes called Boltzmann machines). They can be mainly
categorized into several classes as follows (Amit 1989).

    Let us suppose $N$ is a fixed integer, $S=\{1,2,\cdots,N\}$. We
take the state space as $\{-1,1\}^S$.

\noindent {\bf Model A: deterministic}

{\it A1 (discrete time, synchronous, McCulloch-Pitts):} Denote the
state of the $n$-th step as $X_n=(X_n^1,X_n^2,\cdots,X_n^N)$, then
the dynamic is
\begin{equation}
    X_{n+1}^i=\textrm{sign}\footnote{The function $\textrm{sign}(x)=\left\{\begin{array}{ll}1&x>0;\\0&x=0;\\-1&x<0.\end{array}\right.$}\left(
    \sum_{j=1}^N T_{ij}X_n^j-U_i\right), ~\textrm{if}~
    \sum_{j=1}^N T_{ij}X_n^j\neq U_i;
\end{equation}
and if $\sum_{j=1}^N T_{ij}X_n^j=U_i$, then $X_{n+1}^i$ randomly
choose $1$ or $-1$ with probability $\frac{1}{2}$ respectively.
$U_i~(1\le i\le N)$, given {\it a priori}, are called the threshold
of the $i$th unit.

{\it A2 (continuous time, synchronous):} Every state has an
exponentially distributed stochastic waiting-time, with mean
waiting-time $\lambda^{-1}$, then chooses the next state by the same
rule of model $A1$.

{\it A3 (discrete time, asynchronous, Hopfield):} The neurons are
updated one by one, in some prescribed sequence, or in a random
order. If the previous state $\sigma$ satisfies $\sum_{j=1}^N
T_{ij}\sigma_j>U_i$, then $\sigma_i$ changes to be $1$, otherwise
changes to be $-1$.

{\it A4 (continuous time, asynchronous):} Every state has an
exponentially distributed stochastic waiting-time, with rate
constant $\lambda$, then chooses the next state by the same rule of
model $A3$.

\begin{rem}
Note that by deterministic, we mean the transition from one state to
the next is deterministic.  But the systems with continuous-time
will still behave stochastically due to the Poisson nature in the
transition time.
\end{rem}

\noindent {\bf Model B: stochastic Boolean networks}

{\it B1 (discrete time, synchronous):} Consider the Markov chain
\begin{equation}
    \{X_n=(X_n^1,X_n^2,\cdots,X_n^N), n = 0,1,2,\cdots\}
\end{equation}
on state space $\{-1,1\}^S$, with transition probability given as
follows: for each pair of states $\sigma,\eta\in \{-1,1\}^S$, the
probability transiting from $\sigma$ to $\eta$
\begin{equation}
  p_{\sigma\eta}=\prod_{i=1}^N \frac{\exp(\beta\eta_i(\sum_{j=1}^N T_{ij}\sigma_j-U_i)}
    {\exp(\beta(\sum_{j=1}^N T_{ij}\sigma_j-U_i))
    +\exp(-\beta(\sum_{j=1}^N T_{ij}\sigma_j-U_i))},
\end{equation}
where $\beta (>0)$, $T_{ij}$ and $U_i~ (1\le i,j\le N)$ are
parameters of the model.

{\it B2 (continuous time, synchronous):} Consider the
continuous-time Markov chain $\{\xi_t:t\geq 0\}$ on state space
$\{-1,1\}^S$.  Every state waits an exponential time with meantime
$\lambda^{-1}$ until choosing the next state by the rule of model
$B1$. So the transition density matrix is
\begin{equation}
    q_{\sigma\eta}=\lambda p_{\sigma\eta},
    ~\forall~\sigma,\eta\in \{-1,1\}^S.
\end{equation}

{\it B3 (discrete time, asynchronous):} Denote  $\sigma^i$ to be the
new state which changes the sign of the $i$th coordinate of
$\sigma$. Consider the Markov chain
$\{X_n=(X_n^1,X_n^2,\cdots,X_n^N), ~n=0,1,2,\cdots\}$ on state space
$\{-1,1\}^S$. The neurons are updated one by one, in some prescribed
sequence, or in a random order. Then choose the next state according
to the probability:
$$p_{\sigma\sigma^i}=\frac{\exp(\beta\sigma_i(\sum_{j=1}^N T_{ij}\sigma_j-U_i))}{\exp(\beta(\sum_{j=1}^N T_{ij}\sigma_j-U_i))+\exp(-\beta(\sum_{j=1}^N T_{ij}\sigma_j-U_i))},~\sigma \in \{-1,1\}^S,$$
where $\beta>0$; and $p_{\sigma\eta}=0$, if $\eta\neq \sigma^i$ for
each $i$.

{\it B4 (continuous time, asynchronous):} Every state waits an
exponential time with meantime $\lambda^{-1}$ until choosing the
next state by the rule of model $B3$. So the transition density
matrix is
\begin{equation}
    q_{\sigma\sigma^i}=\lambda p_{\sigma\sigma^i},~\forall \sigma\in \{-1,1\}^S.
\end{equation}

    The third class given below is a variant of the
model $B1$.  It is included here since it is the model used for the
probabilistic Boolean network of cell-cycle regulation in (Ge et al.
2007).

\noindent {\bf Model C: deformation}

Fix $\alpha>0$. Consider a new Markov chain
$\{X_n=(X_n^1,X_n^2,\cdots,X_n^N), n=0,1,2,\cdots\}$ on the state
space $\{-1,1\}^S$ taking the model $B1$ as defined initially, with
transition probability given as follows:
\begin{equation}
    P(X_{n+1}|X_n)=\prod_{i=1}^N P_i(X_{n+1}^i|X_n),
\end{equation}
where we define
\begin{eqnarray}
&&P_i(X_{n+1}^i|X_n)=\nonumber\\
&&\left\{\begin{array}{ll}\frac{\exp(\beta X_{n+1}^i(\sum_{j=1}^N
T_{ij}X_n^j-U_i))}
    {\exp(\beta(\sum_{j=1}^N T_{ij}X_n^j-U_i))+\exp(-\beta(\sum_{j=1}^N T_{ij}X_n^j-U_i))}
        &\textrm{if}~ \sum_{j=1}^N T_{ij}X_n^j\neq U_i\\
\frac{1}{1+e^{-\alpha}}
        &\textrm{if}~ \sum_{j=1}^N
T_{ij}X_n^j=U_i ~\textrm{and}~
X_{n+1}^i=X_n^i\\\frac{e^{-\alpha}}{1+e^{-\alpha}}
        &\textrm{if}~
\sum_{j=1}^N T_{ij}X_n^j=U_i ~\textrm{and}~ X_{n+1}^i=1-X_n^i.
\end{array}\right.\nonumber
\end{eqnarray}

\begin{rem}
The model C differs from B1 when $\sum_{j=1}^NT_{ij}X_n^j-U_i=0$,
and the latter is also a special case of the former when $\alpha=0$.
\end{rem}

\section{Mathematical Circulation Theory of Network Nonequilibrium Steady
States}

This section is also mainly restated from (Ge et al. 2007), ignoring
those detailed proofs.

There are many different approaches to the theory of nonequilibrium
statistical mechanics in the past (Nicolis et al. 1977, Keizer
1987), mathematical theories of which have emerged in the last two
decades, and Jiang et.al. have summarized their results of this
theory in a recent monograph (Jiang et al. 2004). The most important
concepts in the theory are ($i$) reversibility of a stationary
process that corresponds to thermodynamic equilibrium, and ($ii$)
the circulation in a stationary process which corresponds to NESS. A
key result of the theory is the circulation decomposition of NESS.

\subsection{Circulation theory of nonequilibrium steady states}

Hill (Hill 1989) constructed a theoretical framework for discussions
of vivid metabolic systems, such as active transport, muscle
contractions, etc. The basic method of his framework is diagram
calculation for the cycle flux on the metabolic cycles of those
systems (Hill 1989). He successively found that the result from
diagram calculation agrees with the data of the numbers of
completing different cycles given by random test (Monte Carlo test),
but did not yet prove that the former is just the circulation rate
in the sense of trajectory of a corresponding Markov chain. In
Chapter 1 and 2 of (Jiang et al. 2004), Markov chains with discrete
time and continuous time parameter are used as models of Hill's
theory on circulation in biochemical systems. The circulation rate
is defined in the sense of trajectories and the expressions of
circulation rate are calculated which coincide with Hill's result
obtained from diagrams. Hence the authors verify that Hill's cycle
flux is equivalent to the circulation rate defined in the sense of
trajectories.

Below we only state the circulation theory of discrete-time Markov
chains, and refer to Chapter 2 in (Jiang et al. 2004) for the quite
similar circulation theory of continuous-time Markov chains.

First of all, we state the main results, rewritten in terms of the
cycle representation of stationary homogeneous Markov chains
(Kalpazidou 1995, Theorem 1.3.1), which is analogous to the
Kirchoff's current law and circuit theory in the networks of master
equation systems (Schnakenberg 1976).

\begin{thm}
Given a finite oriented graph $G=(V,E)$ and the weight on every edge
$\{w_e>0: e\in E\}$. If the weight satisfies the balance
equation(input=output)
\begin{equation}\label{balance-eq}
\sum_{e^+=i} w_e=\sum_{e^-=i} w_e,~\forall i\in V,
\end{equation}
then there exist a positive function defined on the oriented cycles
$\{w_c: c=(e_1,e_2,\cdots,e_k), k\in Z^+\}$, such that
\begin{equation}
w_e=\sum_{e\in c} w_c,~\forall e\in E.
\end{equation}
\end{thm}

For a finite stationary Markov chain $X$ with finite state space
$S=\{1,2,\cdots,N\}$, transition matrix $P=\{p_{ij}:~ i,j\in S\}$
and invariant distribution $\bar{\pi}=\{\pi_1,\pi_2,\cdots,\pi_N\}$,
one can take its state space to be the group of vertexes, and
$E=\{e: e^-=i,e^+=j,p_{ij}>0\}$ with weight $\{w_e=\pi_ip_{ij}\}$.
Then notice that (\ref{balance-eq}) is satisfied, since
$\sum_{j}p_{ij}=1$  for each $i$ and
$$\sum_{j}\pi_ip_{ij}=\pi_i=\sum_j \pi_jp_{ji},~\forall i\in V,$$
we can conclude that

\begin{cor} (cycle decomposition)
\label{cycdecomp} For an arbitrary finite stationary Markov chain,
there exists a positive function defined on the group of oriented
circuits $\{w_c: c=(i_1,i_2,\cdots,i_k), k\in Z^+\}$ such that
\begin{equation}
\pi_ip_{ij}=\sum_{c} w_cJ_c(i,j),~\forall i,j\in V,
\end{equation}
where $J_c(i,j)$ is defined to be 1 if the cycle c includes the path
$i\rightarrow j$, otherwise 0.
\end{cor}

\begin{defn}
$w_c$ is called the {\bf circulation} along cycle c.
\end{defn}

For any $i,j\in S, i\not=j$,
\begin{equation}
 \pi_i p_{ij}-\pi_j p_{ji}
  =\sum_{c} (w_c-w_{c_-}) J_c(i,j),
 \label{circ-decomp-con}
\end{equation}
where $c_-$ denotes the reversed cycle of $c$. Equation
(\ref{circ-decomp-con}) is called the {\bf circulation
decomposition} of the stationary Markov chain $X$, and the
difference between the circulations of positive direction and
negative direction along the cycle c (i.e. $w_c-w_{c_-}$) is called
the {\bf net circulation} of cycle c, which actually represents the
true flux of this cycle.

It can be proved that generally the circulation decomposition is not
unique, i.e. it is possible to find another set of cycles ${\cal C}$
and weights on these cycles $\{w_c|c\in {\cal C} \}$ which also fit
(\ref{circ-decomp-con}).

However, the most reasonable choice of circulation definition is the
one defined in the sense of trajectories form the probabilistic
point of view. Along almost every sample path, the Markov chain
generates an infinite sequence of cycles, and if we discard every
cycle when it is completed and at the meantime record it down, then
we can count the number of times that a specific cycle $c$ is formed
by time $t$, which we denote by $w_{c,t}(\omega)$.

The following theorem is recapitulated from Theorem 1.3.3 in (Jiang
et al. 2004).
\begin{thm}\label{Thm-circulation}
Let ${\cal C}_n(\omega)$, $n=0,1,2,\cdots$, be the class of all
cycles occurring until $n$ along the sample path $\{X_l(\omega)\}$.
Then the sequence $({\cal C}_n(\omega),w_{c,n}(\omega)/n)$ of sample
weighted cycles associated with the chain $X$ converges almost
surely to a class $({\cal C}_{\infty},w_c)$, that is,
\begin{equation}
 {\cal C}_{\infty}=\lim_{n\rightarrow +\infty} {\cal C}_n(\omega), {\rm ~~a.e.}
\end{equation}
\begin{equation}
 w_c=\lim_{n\rightarrow +\infty} \frac{w_{c,n}(\omega)}{n}, {\rm ~~a.e.}
\end{equation}
Furthermore, for any directed cycle
 $c=(i_1,i_2,\cdots,i_s)\in{\cal C}_{\infty}$, the weight $w_c$ is given by
\begin{equation}
 w_c=p_{i_1i_2}p_{i_2i_3}\cdots p_{i_{s-1}i_s}p_{i_si_1}
      \frac{D(\{i_1,i_2,\cdots,i_s\}^c)}
           {\sum_{j\in S} D(\{j\}^c)}.
\end{equation}
where $D=\{d_{ij}\}=I-P=\{\delta_{ij}-p_{ij}\}$ and $D(H)$ denotes
the determinant of D with rows and columns indexed in the index set
H. The function $\delta_{ij}=\left\{\begin{array}{ll}0,&i\neq
j;\\1,&i=j,\end{array}\right.$ is the well known Kronecker delta
function.
\end{thm}

It is important to emphasize that the circulations defined in the
above theorem also satisfy the circulation decomposition relation
(\ref{circ-decomp-con}).

The above theorem not only rigorously confirms the Hill's theory,
but also gives a prior substitute method of the widely used
diagrammatic method. The complexity of directed diagrams and cycles
increases rapidly with the number of states in the model, while the
determinant interpretation is much more systematic and easy to be
applied using the mathematics softwares. But it will still cost
excessive time to compute these determinants if there are hundreds
of states in the model.

Luckily, a Monte Carlo method using the so-called {\bf derived chain
method} (Section 1.2 in Jiang et al. 2004) to compute the cycle
fluxes has already been developed according to the above theorem,
the main idea of which is just simply to discard every cycle when it
is completed and at the meantime record it down so as to count the
number of times that a specific cycle $c$ is formed by time $t$(i.e.
$w_{c,t}(\omega)$). Then when the time $t$ is long enough, one gets
the approximated circulation of the cycle $c$ (i.e.
$w_c\approx\frac{w_{c,t}(\omega)}{t}$).

Recently, the Boolean yeast cell-cycle network model discussed in
(Ge et al. 2007) has $2048$ states and it is found that the Monte
Carlo method is much more efficient than the method of determinant
interpretation, because the latter is even impossible to be applied
to such a large model upon a normal computer.

The relationship between circulation and NESS is as follows, which
is recapitulated from Theorem 1.4.8 in (Jiang et al. 2004).
\begin{thm}\label{rev-cir-Kol}
Suppose that $X$ is an irreducible and positive-recurrent stationary
Markov chain with the countable state space $S$, the transition
matrix $P=(p_{ij})_{i,j\in S}$ and the invariant probability
distribution $\Pi=(\pi_i)_{i\in S}$, and let $\{w_c:c\in{\cal
C}_{\infty}\}$ be the circulation distribution of $X$, then the
following statements are
equivalent: \\
(i) The Markov chain $X$ is reversible. \\
(ii) The Markov chain $X$ is in detailed balance, that is,
 $$\pi_i p_{ij}=\pi_j p_{ji},\forall i,j\in S.$$
(iii) The transition probability of $X$ satisfies the {\bf
Kolmogorov cyclic condition}:
 $$p_{i_1i_2}p_{i_2i_3}\cdots p_{i_{s-1}i_s}p_{i_si_1}
   =p_{i_1i_s}p_{i_si_{s-1}}\cdots p_{i_3i_2}p_{i_2i_1},$$
for any directed cycle $c=(i_1,\cdots,i_s)$. \\
(iv) The components of the circulation distribution of $X$ satisfy
the symmetry condition:
 $$w_c=w_{c_-},\forall c\in{\cal C}_{\infty}.$$
\end{thm}

Consequently, when the system is in a nonequilibrium steady state,
there exists at least one cycle, containing at least three states,
round which the circulation rates of one direction and its opposite
direction are asymmetric (unequal), so as to cause a net circulation
of the cycle. In theoretic analysis, if there is  a large separation
in the magnitude of the circulation, between few dominant, main
cycles and the rest, it gives rise to the stochastic synchronization
phenomenon and helps to distinguish the most important main
biological pathways, which can be observed in experiments.

\subsection{Applied to deterministic and stochastic Boolean networks}

The keys to understand synchronization behavior in stochastic models
are ($i$) establishing a correspondence between a stochastic
dynamics and its deterministic counterpart; and ($ii$) identifying
the cyclic motion in the stochastic models.

In the framework of the stochastic theory, deterministic models are
simply the limits of stochastic processes with vanishing noise. This
is best illustrated in the following proposition.

\begin{prop}
\label{prop_stod} With the same initial distribution, when
$\beta\rightarrow\infty$, the model $B_k$ converges to the model
$A_k$ in distribution, for $k=1,2,3,4$.
\end{prop}

  We shall further discuss the necessary condition for stochastic
Boolean networks to have cyclic motion. In the theory of neural
networks, one of Hopfield's key results (Hopfield 1982,1984) is that
for symmetric matrix $T_{ij}$, the network has an energy function.
In the theory of Markov processes, having a potential
function(Kolmogorov cyclic condition in Theorem \ref{rev-cir-Kol})
is a sufficient and necessary condition for a reversible process. A
connection is established in the following theorem.

\begin{thm}\label{thm-ness}
For $k=1,2,3,4$, the Markov chain $\{X_n\}$ in the model $B_k$ is
reversible if and only if $T_{ij}=T_{ji}, \forall i,
j=1,2,\cdots,N$, i.e. the matrix $T$ is symmetric.
\end{thm}

From the above two conclusions, it is obvious that
\begin{cor}\label{Cor1}
When $T_{ij}=T_{ji}, \forall i, j=1,2,\cdots,N$, there doesn't exist
any limit cycle consist of more than two states in model $A_k$,
$k=1,2,3,4$.
\end{cor}

\begin{lem}\label{lem}
For $k=1,2,3,4$, the Markov chain $\{X_n\}$ in the model $B_k$ is
reversible when $\beta$ is zero, and the Markov chain $\{X_n\}$ in
the model $C$ is reversible when $\beta$ and $\alpha$ are both zero.
\end{lem}